\documentstyle[bibnorm]{lamuphys}
\newcommand{\be}{\begin{equation}}
\newcommand{\ee}{\end{equation}}

\begin{document}
\title{The M five brane on a torus}

\author{David\,Berman\inst{1}}

\institute{Institute for Theoretical Physics, Princetonplein 5,
3588CC Utrecht, Netherlands}

\maketitle

\begin{abstract}

The D-3 brane is examined form the point of view of the wrapped M-theory
five brane on a torus. In particular, the S-dual versions of the 3-brane
are identified as coming from the different gauge choices of the
auxiliary field that is introduced in the PST description of the five
brane world volume theory.

\end{abstract}
\section{Introduction}

The world volume of the M-theory five brane supports a self dual two
form potential. There are well known difficulties in writing down an
action formulation for such a field. In certain circumstances however
it is possible to write down a covariant action for such a field. The
self duality condition is realized through the introduction of an
auxiliary field \cite{pst3,pst5,pst2}. The role of this field is to
introduce a new gauge symmetry that makes half the degrees of freedom
associated with the two form to be pure gauge. Gauge fixing then
eliminates the unwanted anti self dual degrees of freedom.

There is a conjectured SL(2,Z) duality in IIB string theory \cite{13}. Under the
SL(2,Z) transformation, the Neveu Schwarz and Ramond two forms transform
as a doublet while the axion-dilaton undergoes a fractional linear
transformation. The self dual Ramond Ramond four form however is inert
under this transformation. This is the potential that couples to the D3
brane. Thus a D3 brane is mapped to itself. The world volume theory of the
D3 brane is a Dirac Born-Infeld theory. Such a theory has been shown to
exhibit SL(2,Z) duality properties. As (in the Einstein frame) the complex
coupling on the three brane world volume is the string coupling
(complexified with the Ramond Ramond scalar) the duality properties of the
3-brane world volume theory may be seen as been induced from the duality
properties of the IIB string theory.

IIB string theory reduced on a circle is T-dual to M theory on a torus \cite{12,12.1}.
This is most easily seen at the level of the supergravity actions. In nine
dimensions one may identify the D=11 SUGRA reduced on a torus with the
D=10 IIB SUGRA reduced on a circle \cite{berg}. The SL(2,Z) symmetry of the IIB theory
then becomes associated with the modular group of the torus.

In order to complete the identification one must determine how the branes
of the theory become identified. The D3 brane not wrapped on the circle
will be identified with the five brane wrapped on the torus. The structure
of the paper will be as follows. First we will describe the five brane
world volume action and its reduction on a torus. As part of the reduction
we will discuss the gauge fixing the PST symmetry. The D3 brane and its
reduction will then be described. Then in order to identify the two
theories it will be necessary to dualize the scalar field on the three
brane associated with transverse oscillations in the direction of the
compact dimension.

The moduli and fields of the theories will then be identified and shown to
be in agreement with supergravity and string/membrane calculations. In
particular the role of the self dual field and its realization in the PST
formulation will be examined with respect to the SL(2,Z) transformation of
the D3 brane.

\section{The M-theory five brane}

We work
with
a flat Minkowski background, using a metric, $\eta=diag(-1,+1,+1,..)$.
The $X^M$ are
11-dimensional
space-time coordinates ($M,N = 0..9,11$). $\sigma^{\hat \mu}$ are the
coordinates of the
brane.
The action will also contain
a world volume self
dual two form gauge field, $B$ whose field strength is as usual given by
$H=dB$.

The action
for the 5-brane will be written as follows, \cite{schwarz4,pst2}: \be S=-\int_{M^6} d^6\sigma   
\sqrt{-det \Bigl(G_{\hat \mu \hat \nu} + i{ {\tilde
{\cal{H}}}_{\hat \mu
\hat \nu} \over{
\sqrt{ v^{\hat \mu} v_{\hat \mu}} }} \Bigr)} - {{ {\sqrt{-G}} {\tilde   
{\cal{H}}}^{\hat \mu \hat \nu}
H_{\hat \mu \hat \nu \hat \rho} v^{\hat \rho} } \over { 4 v^2} } 
\label{5.1} \ee where: \be {\tilde{\cal{H}}}_{\hat \mu \hat \nu}
= {1 \over 6}   G_{\hat \mu \hat \alpha} G_{\hat \nu \hat \beta}
{ {\epsilon^{\hat \alpha
\hat \beta \hat \delta \hat \gamma \hat \rho \hat \sigma}} \over
{\sqrt{-G}} } H_{\hat \delta \hat \gamma \hat \rho} v_{\hat \sigma}
  \label{5.2} \ee and \be G_{\hat \mu \hat \nu}=
dX^M_{\hat \mu} dX^N_{\hat \nu} \eta_{MN} \label{5.3} \ee
$G=detG_{\hat \mu \hat \nu}$; $v$ is a completely
auxiliary closed one form field introduced to allow the self-duality
condition to be imposed
in
the action while maintaining Lorentz invariance. Usually the action
\ref{5.1} is
written with $v=da$; however this is only locally correct
as v is constrained to be closed but not necessarily exact.
See the
references \cite{pst5}
for a discussion on this Lorentz invariant formulation.
 Apart from the usual gauge symmetries associated with
the
gauge
potential $B$ and the background field $C$, this action has additional
local, so called {\it{PST}}
symmetries one of which we will use later to eliminate half the degrees 
of freedom of the two form gauge field. \be \delta B =
\chi \wedge v \label{5.5} \ee
 
This will be the action that we will double
dimensionally reduce on
$T^2$. And so we send, $M^6 \rightarrow M^4 \times T^2$
and $M^{11} \rightarrow M^9
\times T^2$. We will identify
\be (X^{11},X^9)=(\sigma^4 , \sigma^5)= (y^1,y^2) \ee Where $(y^1,y^2)$
are
the
coordinates
on
the space-time torus. In these coordinates we will identify $y^1= y^1+1$
and $y^2=y^2+1$. We will drop all functional dependence of the fields on the compact
coordinates, that is taking only the zero modes. $m,n=0..8$ will be the
non compact space-time indices,
$i,j=1,2$ will be torus coordinate indices and $\mu,\nu=0..3$ will be
the
coordinates of the non-wrapped 5-brane world volume. The space-time
metric will be written as \be \eta_{MN} \rightarrow \eta_{mn} \oplus
\eta_{ij} \label{5.6} \ee This truncates the space-time Kaluza Klein fields
associated
with the torus. This is because we are only interested in the M-5
brane/D-3 relationship. Such Kaluza-Klein fields in the M-theory 
picture are associated with the
wrapped D and fundamental string in IIB. We will take $\eta_{mn}$
to be flat Minkowski
metric and take the metric on the torus to be given by \be \eta_{ij}dy^i 
\otimes dy^j= {V \over \tau_2} (dy^1 \otimes 
dy^1 + \tau_1 dy^2 \otimes dy^1+ \tau_1 dy^1 \otimes dy^2 + |\tau|^2 dy^2
\otimes dy^2) \label{5.7}  \ee  $\tau=\tau_1+i\tau_2$ is the complex
structure
of the torus and $V$ is the
area of the torus. The reduction of the brane metric $G$ from \ref{5.3}
follows.

\noindent Similarly, we reduce the world volume gauge field as
follows: \be B=
B_{(0)} dy^1 \wedge dy^2 + B_{(1) i} \wedge dy^i + B_{(2)}  \label{5.12}
\ee so
that
 \be {{H}}= {{J}} +
{{F}}_i \wedge dy^i
+ {{L}} dy^1 \wedge dy^2 \label{513a} \ee

There are two distinct possibilities for the auxiliary one form under this decomposition. One may take $v$ to be in
$T^2$ only or in $M^4$ only. The two choices must be physically
equivalent. The restriction simply corresponds to a partial gauge fixing. 
In what follows we will take
$v$ to be a
member
of the first cohomology on $T^2$. We will consider the
specific choices $v=dy^1$ and $v=dy^2$. These two independent gauge
choices are what will eventually generate the S-duality on the
3-brane. Should we put $v$ in $M^4$, for example $v=dt$ then the SL(2,Z)
symmetry of the
3-brane will become manifest in the action but we will lose manifest
Lorentz invariance. This will give an action of type given in \cite{sands},
\cite{dsb1} The
relationship between the formulation of the
reduced action and the different
gauge choices for the PST one form was discussed in \cite{dsb1,dsb2}

Our goal will be to compare with
the D-3 brane, hence it is natural to express the six dimensional determinant appearing in the action as a four 
dimensional determinant. For this purpose one may use the well known identities:

\begin{eqnarray} det \left( \matrix{ L & P \cr Q & J \cr} \right) = det
\left(\matrix{ L-Q^T
J^{-1} P & 0 \cr 0 & J \cr} \right) \end{eqnarray} and
 \be det(A \oplus B)=det(A)det(B) \ee We have \be detM = det(M_{ij})
det(M_{\mu \nu} - M_{\mu
i}^T
(M^{-1})^{ij} 
M_{j \nu} )  \ee 
This gives the following action:

\be S_{5-2}=-\int_{T^2} \int_{M^4} \sqrt{\eta} \sqrt{-det(G_{\mu \nu} + i 
{\bf{\alpha}}^i(v) {}^* {{F}}_{(i) \mu \nu} - {\bf{\beta}}(v){}^*
{{J}}_\mu 
{}^*{{J}}_\nu )} + {1 \over 2}  {{{F}}}_i \wedge  {{{F}}}_j \gamma^{ij}(v)
\label{5.28} \ee where $\alpha^i(v)$ and $\beta(v)$ and
$\gamma(v)^{ij}$ are constants that
remain to be evaluated and will be dependent on our choice of $v$. 

However, before evaluating them we will put the $\sqrt{\eta}$ inside the
determinant. This becomes $\eta^{1 \over 4}$ inside the determinant. We
will then carry out a Weyl scaling 
so that we absorb this factor into the rescaled 
metric. That is
\be X^\prime = X        \eta^{1/8} \qquad G^\prime_{\mu \nu}= G_{\mu
\nu} \eta^{1 \over 4}  \label{5.29} \ee We then rewrite the action in
this rescaled metric taking care with factors of $\eta$. The $T^2$
integral is trivial.

We will use the symmetry given by equation {\ref{5.5}} to eliminate
half the degrees of freedom contained in the gauge fields. For the
choice $v=dy^L$ we gauge away $F_{(L)}$  and $L_{12}$. This leaves
only one vector gauge field in the action, with field strength $F$,
and one two form gauge field, with field strength $J$. The PST part of
the action will then contribute a total derivative that we shall be
able to identify it with an axion coupling.   We will now write the
action in its final form as follows:

 \be S_{5-2}=- \int_{M^4} \sqrt{-det(G^\prime_{\mu \nu} + i
{\bf{\alpha}}(v) {}^* {{F}}_{ \mu \nu} - {\bf{\beta}}{}^* {{J}}_\mu
{}^*{{J}}_\nu )} + {1 \over 2}  {{{F}}}_i  \wedge {{{F}}}_j
\gamma^{ij}(v) \label{5.30} \ee

We now consider the two natural independent gauge choices for $v$ and
evaluate the coefficients, $\alpha$, $\beta$ and $\gamma$.

We now consider the two natural independent gauge choices for $v$ and
evaluate the coefficients, $\alpha$, $\beta$ and $\gamma$.

\smallskip
\noindent For $v=dy^1$:

\be \alpha= \sqrt{\tau_2 \over |\tau|^2} \qquad \beta= \eta^{3/4}
\qquad \gamma= -{\tau_1 \over |\tau|^2}
\label{5.31a} \ee 

\smallskip
\noindent for $v=dy^2$: \be \alpha= \sqrt{\tau_2} \qquad \beta=
\eta^{3/4} \qquad \gamma=\tau_1
\label{5.31b} \ee 
\smallskip
\noindent

Note that the vector fields couple only to the complex structure of
the torus. That is the couplings are completely determined by the
shape of the torus and are independent of its size. Different choices
of $v$ give different couplings. The opposite is true for the two form
fields. The coupling for the two form field is independent of the
choice of $v$ and is dependent only on the area of the torus. Note
that this is frame dependent statement that is reliant on the Weyl
rescaling.   Combining $\tau= \tau_1 +i \tau_2$ we see the different
choices of $v$ generate the transformation $\tau \rightarrow {-1 \over
\tau}$ in the vector field couplings. This corresponds to one of the
generators of SL(2,Z) the modular group of the torus. The other
generator will arise from an integral shift in $\tau_1$ which will
cause a trivial shift in the total derivative term. Later when we
compare with the 3 brane on $S^1$, we will identify the complex
structure of the torus with the axion-dilaton and the area of the
torus will be related to the radius of the compact dimension as given
in \cite{12,12.1}.

{\section{The D-3 brane}}

Starting with the 10 dimensional IIB three brane action in 10
dimensions \cite{schwarz4}, we will directly reduce the action
on a circle.  The action in the Einstien frame, including an axion
coupling, is given by:

\be S_3=-\int d^4\sigma \sqrt{-det(G_{\mu \nu}+ e^{- {\phi \over 2}}
{{F}}_{\mu \nu} ) } + {1 \over 2} C_0 F \wedge F       \label{5.32} \ee

We will reduce this action directly implying we will not identify any
of the brane coordinates with the compact dimension. Hence, we will
write $X^9=X^9+1 = \phi$ and so decompose the background metric
$g_{mn}\rightarrow g_{mn} \oplus R^2$ where R is the circumference of
the compact dimension. That is as before we truncate out the space
time Kaluza Klein field. (On the M-theory side this corresponds to
truncating the wrapped membrane).

This gives for the induced world volume metric, pulling out the dependence on $\phi$: \be G_{\mu \nu}
\rightarrow G_{\mu \nu} + R^2 \partial_\mu \phi \partial_\nu \phi
\label{35} \ee The world volume gauge field is left invariant.

So the final reduced action for the three brane becomes:

\begin{eqnarray}  
S_{3 , (S^1)}& =&-\int d^4 \sigma \sqrt{-det(G_{\mu \nu} + e^{-{\phi
\over 2}} {{F}}_{\mu \nu} + R^2 \partial_{\mu}    \phi  \partial_\nu
\phi ) } + {1 \over 2} C_0 F \wedge F    \label{5.37} \end{eqnarray}

We wish to compare the wrapped 5-brane with different choices of $v$
with the 3-brane and its S-dual. The S-dual 3-brane is determined by
dualizing the vector field on the brane using the same method as
described below for dualizing the scalar field.

The action with the vector field dualized takes the same form (this is
not true when the full supersymmetric action is used) but the axion
dilaton, $\lambda= C_0 +i e^{- \phi}$ is inverted in the dual
action. That is the usual, $\lambda \rightarrow {-1 \over \lambda}$,
\cite{schwarz4,tseytlin2}.

In order to exactly identify the  reduced 3-brane action with the
5-brane wrapped action we will first need to do a world volume duality
transformation on the field $\phi$. This is in the spirit of \cite{14}
whereby world volume dual actions are associated with the M-theory
picture of the brane.

We will dualize the scalar field $\phi$ by replacing its field
strength $d\phi$ with $l$ and then adding an additional constraint
term to the action $S_c= H \wedge (d \phi - l)$. $H$ is a lagrange
multiplier ensuring that $l=d \phi$. To find the dual we first find
the equations of motion for $\phi$ and solve.  This implies $dH=0$
which means we may locally write $H=dB$. Then we must find the
equations of motion for $l$ and solve in terms of $H$. We simplify the
problem by working in the frame in which $F$ is in Jordan form with
eigenvalues $f_1$ and $f_2$. $l_i$ are the components of $l$ and $h_i$
are the components of the dual of $H$. The equations of motion for $l$
are:  \be  h_1= {-(1+f_2^2) \over {\sqrt{-detM}}} l_1 R^2 \qquad h_2=
{(1+f_2^2) \over {\sqrt{-detM}}} l_2 R^2  \nonumber \ee \be h_3=
{(1-f_1^2) \over {\sqrt{-detM}}} l_3 R^2 \qquad  h_4= {(1-f_1^2) \over
{\sqrt{-detM}}} l_4  R^2  \label{5.40} \ee where \be M_{\mu \nu}=
G_{\mu \nu} + {F}_{\mu \nu} + R^2 l_{\mu}  l_{\nu} \ee We then invert
these equations to solve for $l_i$. The solutions are:  \be  l_1=
{(f_1^2-1) \over {\sqrt{-det{\tilde{M}}}}} {h_1 \over R^2}\qquad l_2=
{-(f_1^2-1) \over {\sqrt{-det{\tilde{M}}}}} {h_2 \over R^2} \nonumber
\ee \be l_3= {(1+f_2^2) \over {\sqrt{-det{\tilde{M}}}}} {h_3 \over
R^2} \qquad l_4= {(1+f_2^2) \over {\sqrt{-det{\tilde{M}}}}} {h_4 \over
R^2}    \label{5.41} \ee Where \be {\tilde M}_{\mu \nu} =G_{\mu \nu} +
i {}^*{F}_{\mu\nu}  - {1 \over R^2} ({}^* H)_{\mu}   ({}^* H)_{\nu} \ee

When we substitute these equations into the action we find,
reinstating dilaton dependence and the axion term: \be S_{D}=-\int d^4
\sigma \sqrt{-det\Bigl(G_{\mu \nu} + i e^{-{\phi \over 2}}
{}^*{F}_{\mu \nu}  -  {1 \over R^2} ({}^* H)_{\mu}   ({}^* H)_{\nu}
\Bigr)}  + {1 \over 2} C_0 F \wedge F   \label{5.42} \ee

We wish to compare with the usual M-theory predictions given in
\cite{12} concerning the relationship between the moduli of the IIB
theory in 9 dimensions with the geometrical properties of the torus
used in the M-theory compactification. By comparing the actions for
the reduced five brane and the reduced/dualized three brane on
identifies the moduli in the different theories.  The scaling of the
metric given in equation \ref{5.29} implies \be G^B_{\mu \nu} =
Area(T^2) ^{1 \over 2} G^M_{\mu \nu} \label{5.43} \ee From both the
coefficient in front of F in the determinant and the coefficient in
front of the $F \wedge F$ term, we identify the axion-dilaton of the
IIB theory (in the 10 dimensional Einstein frame) with the complex
structure of the torus. \be \lambda= C_0 +i e^{-\phi} = \tau
\label{5.44} \ee From comparing the coefficient in front of $^*H$, the
radius of the the 10th  dimension in IIB becomes: \be R_B = Area(T^2)
^{-{3 \over4 }}
\label{5.45} \ee We have identified the gauge field on the reduced 5-brane
with the gauge field on the reduced D-3. The dualized scalar on the
D-3 brane becomes identified with the three form on the reduced M-5
brane.
\bigskip

\section{Conclusions/discussion}

The identification of the moduli given above are in agreement with
\cite{12,12.1} as one would expect. The elegant aspect is how the S-dual
versions of the three brane appear from different gauge choices of the
auxiliary field in the five brane. In the discussion above no attempt
has been made to discuss the full supersymmetric versions of the
actions used. This has been explored in detail in \cite{dsb2}. There it
was shown that the conclusions of the above are unaffected by the
presence of Fermions though the dualization procedure becomes a great
deal more involved due to the presence of extra interaction terms that
make the dual action take on a different form. Nevertheless the
different gauge choices of the auxiliary field still allow one to
generate the different dual versions once the correct field redefinitions are
made. See \cite{dsb2} for the details.

Another aspect that has not been explored here is the validity of the
identification at the quantum level. As is known, global
considerations make the construction of partition functions for chiral
gauge fields problematical \cite{wit1,wit2}. Given that classically
one identifies the coupling on the 3 brane with the modular parameter
of the torus, this suggests that the duality properties of the 3-brane
partition function reported in \cite{wit3} are manifest as modular
properties of the 5-brane partition function. Related ideas have been
discussed in \cite{eric}.

%
%

\end{document}